\documentclass[pdflatex,sn-mathphys-num]{sn-jnl}

\usepackage{graphicx}%
\usepackage{multirow}%
\usepackage{amsmath,amssymb,amsfonts}%
\usepackage{amsthm}%
\usepackage{mathrsfs}%
\usepackage[title]{appendix}%
\usepackage{xcolor}%
\usepackage{textcomp}%
\usepackage{manyfoot}%
\usepackage{booktabs}%
\usepackage{algorithm}%
\usepackage{algorithmicx}%
\usepackage{algpseudocode}%
\usepackage{listings}%
\usepackage[utf8]{luainputenc} 
\usepackage[T1]{fontenc}
\usepackage[polish, english]{babel}
\usepackage{xcolor}
\usepackage{placeins}
\usepackage{makecell}
\usepackage{soul}

\usepackage{caption}
\usepackage{ragged2e}

\theoremstyle{thmstyleone}%
\theoremstyle{thmstyletwo}%

\theoremstyle{thmstylethree}%

\begin{document}

\title[]{Prospects for detecting charged long-lived BSM particles at MoEDAL-MAPP experiment: \\
A mini-review}


\author*[1]{\fnm{Rafał} \sur{Masełek}}\email{rafal.maselek@ijs.si}

\author[2]{\fnm{Kazuki} \sur{Sakurai}}\email{kazuki.sakurai@fuw.edu.pl}

\affil*[1]{\orgdiv{Theoretical Physics Department}, \orgname{Jožef Stefan Institute}, \orgaddress{\street{Jamova cesta 39}, \city{Ljubljana}, \postcode{1000},  \country{Slovenia}}}

\affil[2]{\orgdiv{Institute of Theoretical Physics, Faculty of Physics}, \orgname{University of Warsaw}, \orgaddress{\street{Pasteura 5}, \city{Warsaw}, \postcode{02-093}, \country{Poland}}}


\abstract{
The search for physics beyond the Standard Model at the Large Hadron Collider is expanding to include unconventional signatures such as long-lived particles. This mini-review assesses the prospects for detecting electrically charged long-lived particles using the MoEDAL-MAPP experiment.
We synthesize findings from recent studies that evaluate sensitivity to supersymmetric models, radiative neutrino mass scenarios, and generic multiply charged objects. A key component of this review is the comparative analysis of MoEDAL’s reach against the general-purpose ATLAS and CMS experiments. We conclude that while MoEDAL is constrained by lower integrated luminosity, its passive, background-free detection methodology offers a unique advantage. Specifically, the experiment provides complementarity to the major detectors, particularly for signals involving slow-moving particles and stable states with intermediate electric charges.
}


\maketitle

\section{Introduction}\label{sec:intro}

The Standard Model (SM) of particle physics, despite its remarkable success, leaves several fundamental questions unanswered, such as the nature of dark matter, the origin of neutrino masses, and the hierarchy problem. Many theoretical frameworks proposed to address these shortcomings—including Supersymmetry (SUSY), universal extra dimensions, and various hidden sector models—predict the existence of new massive particles. While the majority of searches at the Large Hadron Collider (LHC) focus on prompt decays, a significant subset of these Beyond the Standard Model (BSM) theories naturally gives rise to Long-Lived Particles (LLPs) that can travel appreciable distances through the detectors before decaying~\cite{Alimena:2019zri}.

The search for charged LLPs has become a key frontier at the LHC. The general-purpose detectors, ATLAS and CMS, have deployed extensive search programs targeting Heavy Stable Charged Particles (HSCPs) and multi-charged objects. These analyses typically exploit anomalous ionisation energy loss ($dE/dx$) in the inner trackers and Time-of-Flight (ToF) measurements in the muon systems to identify slow-moving, heavy candidates. 
Recently, the ATLAS collaboration published results using the full Run 2 dataset ($139~\text{fb}^{-1}$), placing stringent limit on meta-stable gluino $R$-hadrons, $m \gtrsim 2$ TeV \cite{ATLAS:2022pib}, as well as multi-charged particles {with electric charges in the range $|q| \in [2e, 7e]$, excluding masses up to $1.6$ TeV for $|q|=6e$ \cite{ATLAS:2023zxo}.
Similarly, the CMS collaboration has updated its HSCP search using Run 2 data ($101~\text{fb}^{-1}$), identifying exclusion limits for doubly charged fermions $\ge 1.4$ TeV \cite{CMS:2024nhn}.
}

Despite these impressive constraints, general-purpose detectors face intrinsic limitations. They rely heavily on specific triggers, often requiring high transverse momentum ($p_{T}$) or missing energy, which can be inefficient for slow-moving ($\beta \ll 1$) or non-standard signal topologies. Furthermore, standard detector electronics may saturate when exposed to the extreme ionization densities produced by highly charged objects ($|q| \gg 1e$), potentially leading to signal loss.

The MoEDAL experiment is able to cover some of these blind spots. MoEDAL employs passive Nuclear Track Detectors (NTDs) and Magnetic Monopole Trappers (MMTs) that are sensitive to highly ionizing particles without the need for electronic triggers or precise timing~\cite{MoEDAL:2009jwa}. This allows for the detection of particles with arbitrarily low velocities and very high electric or magnetic charges, a regime often inaccessible to ATLAS and CMS. Specifically, the MoEDAL collaboration has searched for Highly Electrically Charged Objects (HECOs) with
$|q| \in [15e, 175e]$ 
using Run 1 \cite{MoEDAL:2021mpi} ($E_{\rm CM}=8~{\rm TeV},~L=2.2~{\rm pb}^{-1}$) data, and was able to put mass limits ranging from 70 to 1020 GeV, depending on the electric charge and spin (see Tab. 2 in Ref. \cite{MoEDAL:2021mpi}).
The most recent search was done on Run 2 \cite{MoEDAL:2023ost} ($E_{\rm CM}=13~{\rm TeV},~L=6.46~{\rm fb}^{-1}$) data, where HECOs with charges $q \in [10e, 400e]$ were considered, and the limits spanned from 80 GeV to $\approx$ 3.9 TeV (see Tab. II in Ref. \cite{MoEDAL:2023ost}).

In preparation for LHC Run 3, the experiment has been upgraded to MoEDAL-MAPP. While the new MAPP subdetector extends sensitivity to neutral, feebly interacting particles, the core program for charged LLPs remains centred on the enhanced NTD arrays. This mini-review synthesizes a series of phenomenological studies that evaluate the detector's potential. We discuss discovery prospects for charged LLPs arising from SUSY, neutrino mass models, and generic highly charged scenarios, highlighting the regions of parameter space where MoEDAL-MAPP provides unique, background-free sensitivity complementary to the major LHC experiments.

\section{MoEDAL-MAPP experiment}\label{sec:experiment}

The \textbf{Mo}nopole and \textbf{E}xotics \textbf{D}etector \textbf{a}t the \textbf{L}HC (MoEDAL) is a small, mostly passive LHC detector located about 2 m away from the IP8, outside of 
the LHCb VELO detector. It was designed mainly to discover magnetic monopoles \cite{MoEDAL:2009jwa}, but its search programme 
includes dyons, Q-balls, black hole remnants, and highly ionising massive particles. MoEDAL comprises of 
the three main subdetector systems: Nuclear Track Detectors (NTDs), Magnetic Monopole Trappers (MMTs), 
and TimePix (TPX). 

The main subsystem consists of an array of over a hundred NTD panels, each measuring 
25 cm in width and length, and only a couple of millimetres in depth. Each panel consists of 6 layers of 
special polymer, which degrade when penetrated by a highly ionising particle. NTD is a passive 
detector system; after a long exposure, the NTD array is dismantled and transported to an external laboratory, where 
panels are disassembled, and each polymer layer is etched with a special solvent. The etching dissolves 
the outer layer of the polymer, but along the energy deposition, the speed of dissolving is larger. As a 
result, cone-shaped etch pits are formed on both sides of the NTD panels, revealing the trajectory of 
ionising particles. By analysing the shapes and alignment of etch pits in 6 layers of an NTD panel, the 
presence of a signal can be established. The key properties of this method are: 1) the expected SM 
background is $\ll 1$, making it a background-free experiment; 2) the possibility to reconstruct a track 
depends on the particle's velocity, charge, and incidence angle. The NTD array is the main detector used in searches for charged long-lived particles that are the topic of this article.

The other detector subsystem is MMT, designed to capture magnetic monopoles for later examination, and a pixel detector, TPX, for real-time monitoring of the radiative environment. In preparation for the Run 3 
data taking period, MoEDAL has been augmented with a \textbf{M}oEDAL \textbf{A}pparatus for 
\textbf{P}enetrating \textbf{P}articles (MAPP)\cite{Kalliokoski:2023cgw}, a scintillator detector for discovery of Feebly 
Interacting Particles. The collaboration decided on changing its name and has been using "MoEDAL-MAPP" 
since then.

\section{Methodology of phenomenological studies}\label{sec:methodology}

In the following part of the article, we discuss a series of phenomenological studies 
aimed at assessing the discovery potential for various charged long-lived BSM 
particles at the MoEDAL-MAPP experiment. In contrast to the major LHC experiments, 
such as ATLAS and CMS, no public simulation framework suitable for these signatures 
was available. To address this gap, a dedicated fast-simulation 
framework in \textsc{Python} has been developed. The code focuses on modelling the 
Nuclear Track Detector array of MoEDAL, as it is the only detector subsystem relevant 
for the charged long-lived particles considered in this work.

\begin{figure}[h]
\centering
\includegraphics[width=0.7\textwidth]{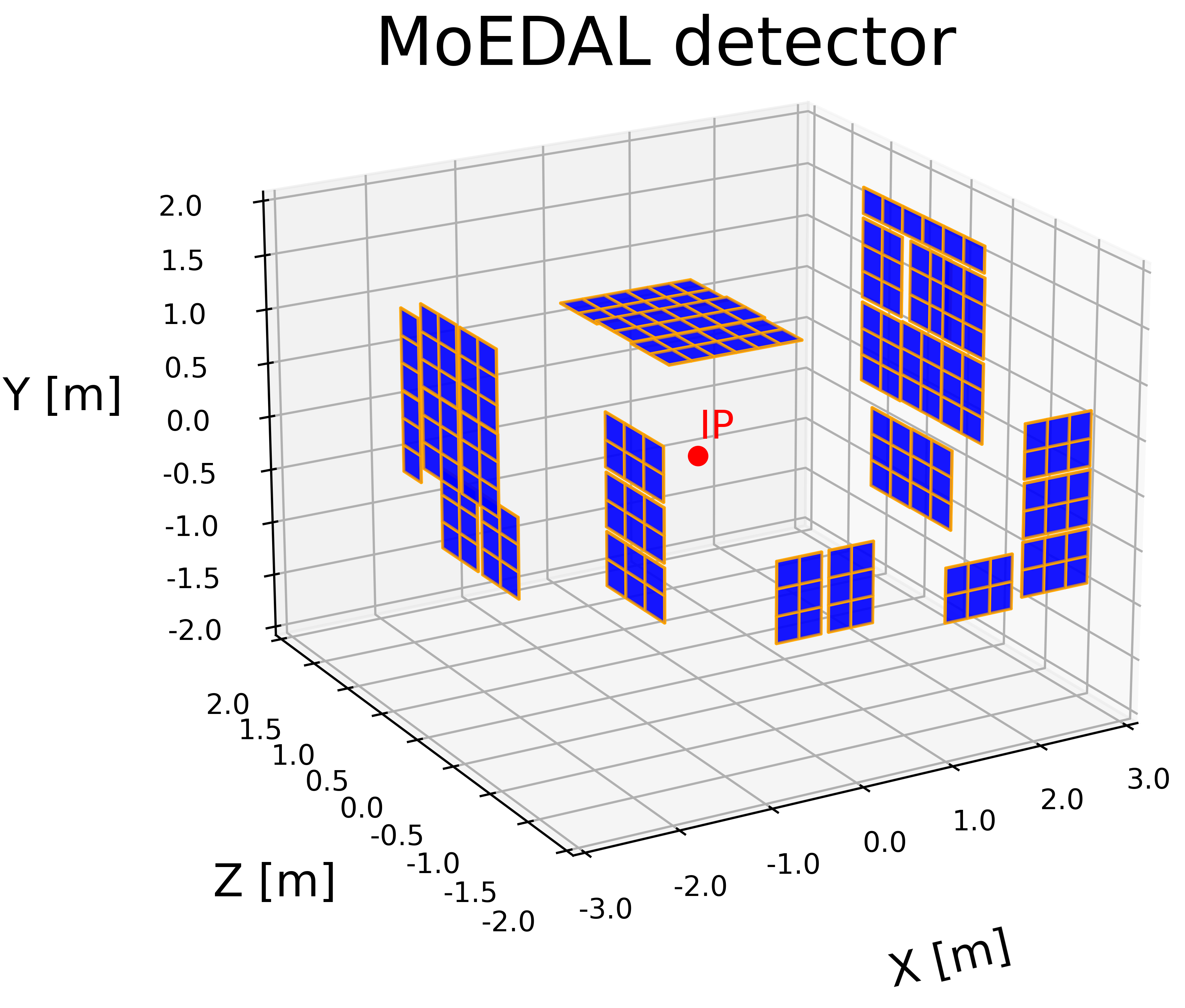}
\caption{\small Nuclear Track Detector panel placement for the MoEDAL Run 2 geometry. Image taken from \cite{Maselek:2023fvy}.
\vspace{-1em}
}\label{fig:methodology:geometry}
\end{figure}

The code accepts \textsc{LHE} files with simulated BSM events as its input. For each 
event, the final state particles are identified, and their four-momenta are 
extracted. Next, the locations and dimensions of the NTD panels are loaded from the 
Run 2 MoEDAL geometry file (see Fig. \ref{fig:methodology:geometry}). Particle 
trajectories are checked for intersecting the NTD array, with a set of simplifying 
assumptions in mind. First, the particles are assumed to propagate between their creation and the detector without 
significant loss of energy or change in momentum. The possibility of a decay is 
handled by assigning weights to particles at a later stage of the analysis. Moreover, 
the depth of the NTD panels is neglected, since it is three orders of magnitude 
smaller than the width and height. Particles whose trajectories do not intersect with 
NTD panels are rejected. The rest are subjected to velocity and incident angle 
constraints, as described in detail in \cite{Maselek:2023fvy}. The partial efficiency of particle $i$ with charge $Q$, three-momentum $\vec{p}_i$, velocity $\beta_i$, incidence angle $\delta_i$, and lifetime $\tau_i$ is given by:
\begin{equation}\label{eq:epsilon-i}
    \epsilon_i = P_{\rm NTD}\left( \vec{p}_i, \tau_i\right) \cdot \Theta 
    \left(
    \delta_{\rm max}\left( \beta_i, Q \right) - \delta_i,
    \right)
\end{equation}
where $P_{\rm NTD}\left( \vec{p}_i, \tau_i\right)$ is a probability of reaching the 
detector, $\Theta$ is the Heavyside step function, and  $\delta_{\rm max}\left( \beta_i, Q \right)$ is the maximum incidence 
angle for particle with velocity $\beta_i$ and charge $Q$, above which it cannot be 
reconstructed. For Run 3, the NTD panels of the MoEDAL detector have been rearranged to face the interaction point, greatly leveraging the efficiency. For such detector placement, the Eq.\ \eqref{eq:epsilon-i} can be approximated with:
\begin{equation}\label{eq:epsilon-i-approx}
    \epsilon_i = P_{\rm NTD}\left( \vec{p}_i, \tau_i\right) \cdot \Theta 
    \left(
    0.15 \cdot |Q|-\beta_i
    \right).
\end{equation}
Eq.\ \eqref{eq:epsilon-i-approx} reveals a crucial property of the MoEDAL detector: 
singly-charged particles have to be slow in order to be detectable at MoEDAL, in 
contrast to major LHC experiments like ATLAS and CMS, which are sensitive only for 
highly boosted particles. However, the velocity threshold grows for particles with 
larger electric charges leading to enhanced sensitivity, in principle, for $|Q|=6\frac{2}{3}$, the velocity constraint becomes irrelevant.

The mean efficiency of the NTD panel array is calculated by summing over all partial efficiencies and averaging over the Monte Carlo sample:
\begin{equation}\label{eq:epsilon-mean}
    \epsilon = 
    \left<
    \sum_{i=1}^N \epsilon_i
    \right>_{\rm MC}
    =
    \left< \sum_{i=1}^N 
    P_{\rm NTD}\left( \vec{p}_i, \tau_i\right) \cdot \Theta 
    \left(
    \delta_{\rm max}\left( \beta_i, Q \right) - \delta_i,
    \right)
    \right>_{\rm MC},
\end{equation}
where we have used $\epsilon_i$ defined by Eq.\ \eqref{eq:epsilon-i}. In Eq.\ \eqref{eq:epsilon-mean}, $N$ is the number of BSM particles in an event, usually $N=2$. The expected number of signal events that MoEDAL will detect over a period of data taking is:
\begin{equation}\label{eq:Nsig}
    N_{\rm sig} = \sigma(m) \times L  \times \epsilon,
\end{equation}
where $\sigma(m)$ is the production cross-section {of charged long-lived particles with mass $m$} and $L$ is the integrated luminosity.

\section{Future discovery prospects}\label{sec:results}


\subsection{Proof-of-concept: neutralinos decaying to meta-stable staus}\label{sec:prospects:neutralino}

Although the possibility of searching for charged long-lived BSM particles at MoEDAL 
was foreseen in the experiment's technical design report \cite{MoEDAL:2009jwa}, the 
first detailed assessment of MoEDAL's sensitivity has been presented 
in Ref. \cite{Felea:2020cvf}. For the needs of this pioneering study, the fast simulation 
framework described in Sec.\ \ref{sec:methodology} was created. In the study, an 
interesting simplified supersymmetric model was introduced. In this model, a pair of 
gluinos is produced, each gluino decays promptly to two quarks and a long-lived neutralino, 
which subsequently decays to an off-shell tau and a metastable stau. The neutralino is assumed 
to be 30 GeV lighter than the gluino, in order to make the jets from gluino decay soft.
A small mass splitting between neutralino and stau is assumed, 1 GeV, to ensure that the 
neutralino is long-lived. 
{Since we are interested in the interaction of a stau with the MoEDAL detector, we assume staus are meta-stable (stable within the detector volume) without specifying its decay modes.
Such suppressed stau deacays can be realised, for example, via small RPV coupling to tau and other SM particles or, in a supergravity scenario, via Planck suppressed coupling to tau and gravitino.}

\begin{figure}[t]
\centering
\includegraphics[width=0.7\textwidth]{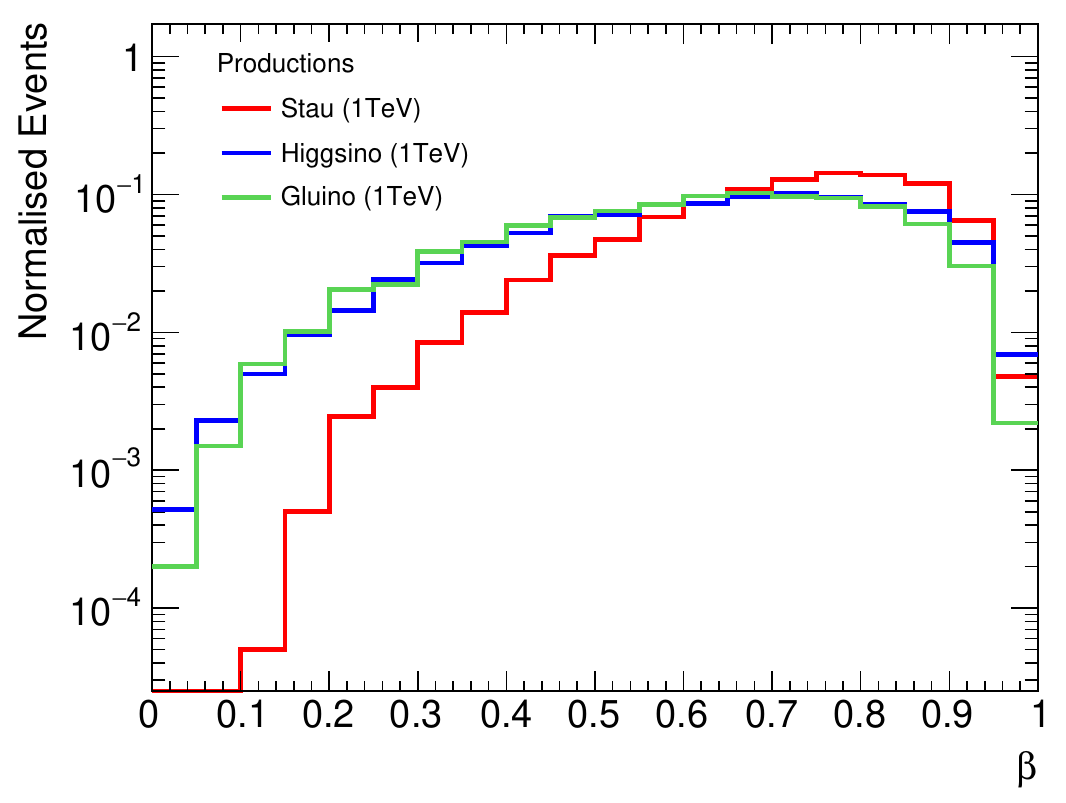}
\caption{\small Velocity distributions for 1 TeV staus, Higgsinos, and gluinos produced via Drell-Yan process. Plot taken from \cite{Felea:2020cvf}.
\vspace{-1em}
}\label{fig:prospects:neutralino:velocity}
\end{figure}

This rather peculiar model 
possesses several features which make it attractive to study in MoEDAL. Firstly, 
gluinos are coloured particles with a relatively large production cross-section. 
Secondly, they are fermions, and fermions produced in the Drell-Yan process do not 
suffer from a p-wave suppression, unlike scalar particles. Thanks to that, their 
velocity distribution is shifted towards smaller values compared to e.g. staus, as 
can be seen in Fig. \ref{fig:prospects:neutralino:velocity}. Thirdly, the neutralino 
is promptly produced but {meta-stable}, leading to decreased efficiency in ATLAS and CMS 
detectors. Its long lifetime is advantageous for MoEDAL, but NTD panels can detect 
only highly ionising particles; therefore, decay to long-lived staus is necessary. 
The proposed model constitutes an optimal scenario from the MoEDAL's point of view. 
It possesses features enhancing MoEDAL's efficiency and reducing the sensitivity of 
major LHC experiments. Therefore, it is a good benchmark to test if MoEDAL can 
compete with ATLAS and CMS. 

For the considered scenario, the probability 
that a stau hits 
an NTD array is given by:
\begin{equation}
    P_{\rm NTD} \left( \vec p_{\tilde \chi_1^0}, \tau_{\tilde \chi_1^0}\right) = 
    \omega \left( \vec p_{\tilde \chi_1^0} \right)
    \left[ 
    1 - \exp\left( 
    \frac{L_{\rm NTD}  \left( \vec p_{\tilde \chi_1^0}\right)}{\beta \gamma c \tau_{\tilde \chi_1^0}}
    \right)
    \right],
\end{equation}
where $ \vec p_{\tilde \chi_1^0}$ and $\tau_{\tilde \chi_1^0}$ are neutralino's momentum and lifetime, respectively, $L_{\rm NTD}$ is the distance from IP to NTD panel along the neutralino's trajectory, $\beta$ is particle's velocity, $\gamma$ is Lorent factor, and 
$ \omega \left( \vec p_{\tilde \chi_1^0} \right)$ is either 1 or 0 depending on whether the particle's trajectory crosses the NTD array or not. 
{The factor in the square-bracket corresponds to the probability that the neutralino decay $\tilde \chi_1^0 \to \tilde \tau \tau^{*}$ occurs before $\tilde \chi_1^0$ reaches an NTD.}
Due to the small mass splitting between the neutralino and the stau, we assume the momenta of the two particles are very similar.

In the study, we consider two scenarios. The first one is to take the Run 2 geometry and 
calculate efficiencies with Eq.\ \eqref{eq:epsilon-i}. The second scenario, which we label 
"ideal", is to assume a slight rearrangement of NTD panels such that they face the IP 
directly. This allows us to neglect the impact of the incidence angle and use an approximation in Eq.\ \eqref{eq:epsilon-i-approx}.

\begin{figure}[t]
\centering
\includegraphics[width=0.7\textwidth]{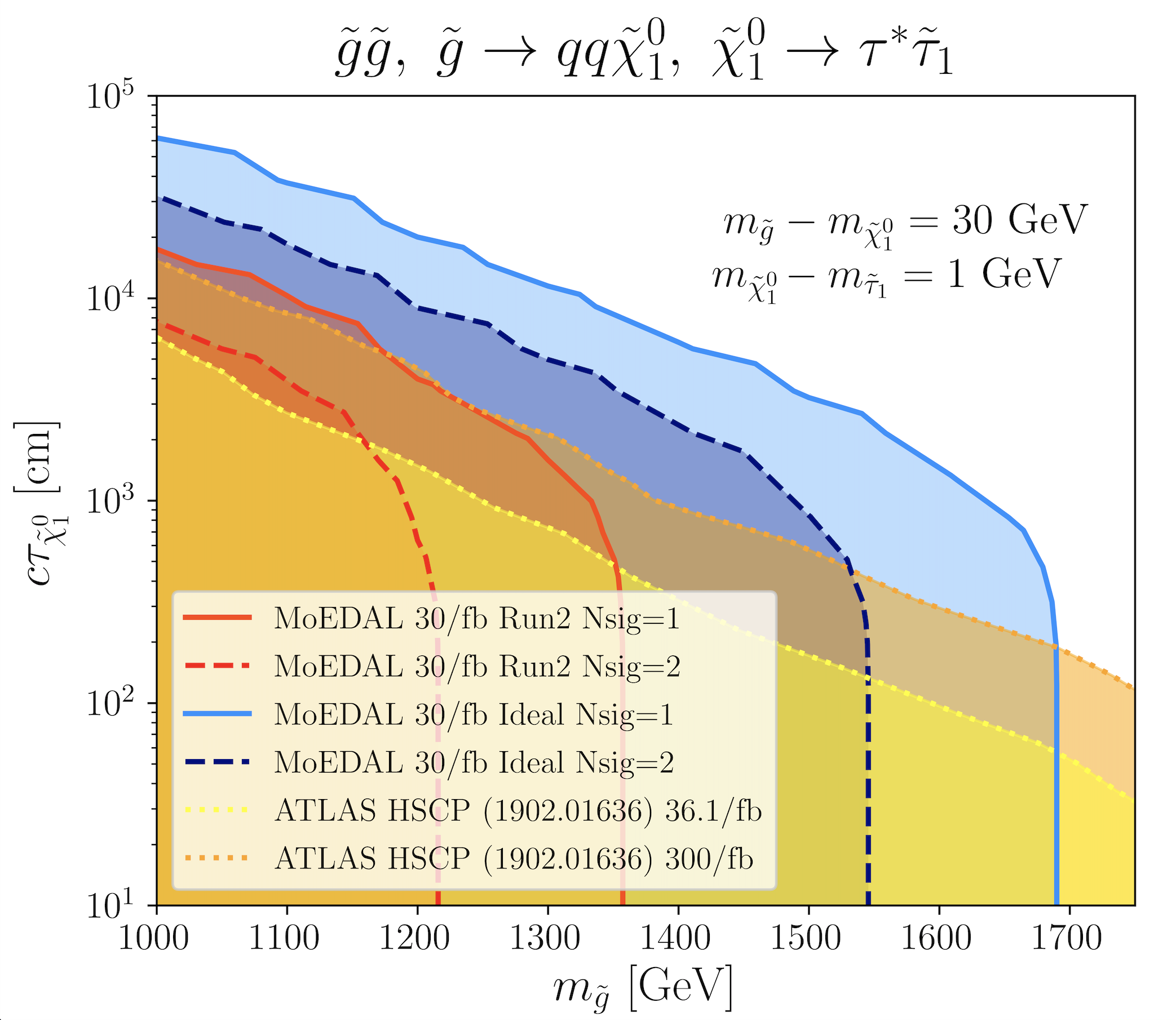}
\caption{\small Expected sensitivity of the MoEDAL experiment to supersymmetric model. Red and blue contours correspond to Run 2 and "ideal" MoEDAL geometries, respectively.
Solid line marks contours for $\rm N_{sig}=1$, while dashed lines are for $\rm N_{sig}=2$. The yellow contour shows constraints from ATLAS search for HSCPs \cite{ATLAS:2019gqq}, done for $L=36.1~{\rm fb}^{-1}$. The orange contour is ATLAS constraint projected for the end of the Run 3 data taking period, $L=300~{\rm fb}^{-1}$. Plot taken from \cite{Felea:2020cvf}.
\vspace{-1em}
}\label{fig:prospects:neutralino:limit}
\end{figure}

The expected sensitivity of MoEDAL at the end of Run 3 ($L=30~\rm{fb}^{-1}$) for the 
considered SUSY scenario is depicted in Fig. \ref{fig:prospects:neutralino:limit}, where 
exclusion contours on gluino mass and neutralino's $c\tau$ are shown, assuming detection of 
$\rm N_{sig}=1$ (solid) or $\rm N_{sig}=2$ (dashed). The exclusion achievable with Run 2 geometry is depicted in red, while the sensitivity of the rearranged detector placement is 
shown in blue. One can see that the "ideal" geometry offers much better sensitivity, allowing 
for testing gluino masses up to 1.7 (1.55) TeV compared to 1.35 (1.2) TeV for Run 2 geometry 
when $\rm N_{sig}=1$ ($\rm N_{sig}=2$). In order to provide a comparison with major LHC 
experiments, we recast the ATLAS search for Heavy Stable Charged Objects \cite{ATLAS:2019gqq}, and depict 
it in Fig. \ref{fig:prospects:neutralino:limit} as a yellow contour. We scale the ATLAS constraint up to Run 3 luminosity ($L=300~\rm{fb}^{-1}$) and depict it in orange. Interestingly, the character of the ATLAS 
constraint is different from MoEDAL's, and the latter offers better sensitivity for a broad 
range of neutralino decay lengths, when the "ideal" geometry set-up is used. This means that 
MoEDAL is able to provide not only independent, but also competitive constraints on some of 
the BSM Physics models. Results of this study lead to MoEDAL-MAPP collaboration's decision to rearrange NTD panels for Run 3, to match the "ideal" set-up.

\subsection{Searches for supersymmetric long-lived particles}\label{sec:prospects:susy}

Encouraged by the promising results described in Sec.\ \ref{sec:prospects:neutralino}, a 
comprehensive evaluation of MoEDAL's performance for more typical BSM scenarios has been 
carried out in Ref. \cite{Acharya:2020uwc}. In this study, the pair production of supersymmetric 
particles was investigated, i.e. long-lived staus, charginos (wino- and Higgsino-like), and R-hadrons formed from 
gluinos, stops or light-flavour squarks. 

A simple phenomenological description of R-hadron 
formation was adopted, assuming that R-hadrons are formed by supersymmetric particles together 
with up or down quarks. The possible bound states involving gluino are $\tilde g  u \bar u$, $\tilde g d \bar d$, $\tilde g u \bar d$, and $\tilde g d \bar u$. The former two bound states 
are electrically neutral, hence they cannot be detected by MoEDAL. We introduce a free 
parameter, $\kappa$, representing a fraction of charged states amongst all produced R-hadrons, and vary it between 0.3 and 0.7. 

For squarks, we consider two scenarios. First, we investigate the possibility when stops are 
relatively light compared to other squark flavours. In the second scenario, stops are heavy, 
and the other five flavours are nearly mass degenerate. In both cases, we assume that heavy 
squarks decay to the lightest one on a time scale $\ll \mathcal{O}(1~\rm m/c) $. The lightest 
squark forms bound states with u and d quarks: $\tilde u \bar u$, $\tilde d \bar d$, $\tilde u \bar d$, $\tilde d \bar u$. Similarly to gluino R-hadrons, only a fraction $\kappa$ of the final states are electrically charged. 

In addition to supersymmetry, we also consider simplified scenarios featuring 
doubly charged particles, motivated by seesaw and left–right symmetric models. Our approach 
relies on a minimal set of assumptions: the particles are pair-produced through the Drell–Yan 
process, carry electric charge $\pm2$e, and are either scalar or spin-1/2 fermionic fields. 
They are singlets under $\rm SU(3)_C$ and transform either as singlets or triplets under $\rm SU(2)_L$.

To calculate the expected sensitivity of MoEDAL at the end of Run 3 ($L=30~\rm{fb}^{-1}$), the 
simulation framework was adjusted to accommodate new types of particles. In 
particular, the probability of reaching the NTD panel was:
\begin{equation}\label{eq:prospects:susy:pntd}
    P_{\rm NTD} \left( \vec p, \tau\right) = 
    \omega \left( \vec p \right)
    \exp\left( 
    \frac{L_{\rm NTD}  ( \vec p)}{\beta \gamma c \tau}
    \right)
   .
\end{equation}
From now on, the effect of the incidence angle is being neglected, which corresponds to the "ideal" geometry set-up described in Sec.\ \ref{sec:prospects:neutralino}.

The results of the study \cite{Acharya:2020uwc}, for $\rm N_{sig}=1$ and $\kappa=0.7$, are 
summarised in Tab.\ \ref{tab:prospects:susy:susy} and Tab.\ \ref{tab:prospects:susy:doubly}, 
together with constraints from ATLAS and CMS. For supersymmetric particles, ATLAS constraints 
for $\tilde g$, $\tilde t$, $\widetilde W$, and $\tilde \tau$ were directly taken from HSCP 
search \cite{ATLAS:2019gqq} with $L=36.1~{\rm fb}^{-1}$. Although ATLAS in \cite{ATLAS:2019gqq} did not interpret its results for light-flavour squarks or Higgsinos, we 
were able to obtain limits on these sparticles by recasting
cross-section upper limits for Winos and sbottoms. The CMS constraints for gluinos, stops, and 
staus come from an analysis \cite{CMS:2016kce} based on a smaller data set, $L=2.5~{\rm fb}^{-1}$. The same analysis provides limits on doubly-charged $\rm SU(2)_L$-singlet spin-1/2 
fermions. With additional recasting, we were able to obtain limits on scalars and fermionic 
triplets.

The results in Tab.\ \ref{tab:prospects:susy:susy} and Tab.\ \ref{tab:prospects:susy:doubly} 
show that, in most scenarios, MoEDAL cannot match the sensitivity achieved by ATLAS or CMS. 
There are, however, two notable exceptions. For the fermionic singlet, MoEDAL and CMS provide 
comparable sensitivity. For the fermionic $\mathrm{SU}(2)_L$ triplet, MoEDAL is expected to 
outperform CMS by roughly 230 GeV. It is important to emphasise, though, that the CMS limits 
were obtained using a relatively small data set of only $L = 2.5~\mathrm{fb}^{-1}$.

MoEDAL’s generally lower sensitivity compared with the major LHC experiments can be traced to 
two main factors. First, 
due to the LHCb experiment's operational constraints, MoEDAL, which shares the same interaction point, 
has access to roughly an order of magnitude less luminosity. 
Second, MoEDAL is only sensitive to slowly moving particles. Most 
particles produced at the LHC are ultrarelativistic unless they are very heavy, but such heavy 
states typically have small production cross sections; the reduced luminosity further 
suppresses MoEDAL’s reach in these cases. The relatively strong performance observed for 
doubly charged fermions arises from their enhanced ionisation power as well as the more 
favourable velocity spectrum of spin-1/2 particles produced via the Drell–Yan mechanism.

\begin{table}[t]
\begin{minipage}[t]{0.49\textwidth}
\caption{\small
Expected MoEDAL's sensitivity to long-lived supersymmetric particles at the end of Run 3 ($L=30~{\rm fb }^{-1}$) compared with ATLAS and CMS constraints (in parentheses). Constraints for squarks and Higgsinos (in double parentheses) were obtained by recasting results from Ref. \cite{ATLAS:2019gqq}.
Table taken from \cite{Acharya:2020uwc}.\\[-5pt]}
\centering
\begin{tabular}{c|c|c|c}
        & MoEDAL & (ATLAS) & (CMS) \\\hline
         $\tilde g$ &  1600 & (2000) & (1500) \\
         $\tilde q$ &  1920 & ((2310)) & - \\
         $\tilde t$ &  920 & (1350) & (1000) \\
         $\widetilde W$ & 670 & (1090) & -\\
         $\tilde h$ & 530 & ((1170)) & -\\
         $\tilde \tau$ & 61 & (430) & (230)
\end{tabular}
\label{tab:prospects:susy:susy}
\end{minipage}
\hfill
\begin{minipage}[t]{0.49\textwidth}
\caption{\small
Expected MoEDAL's sensitivity to long-lived doubly-charged particles at the end of Run 3 ($L=30~{\rm fb }^{-1}$) compared with CMS constraints for singlet fermions (in parentheses) from Ref. \cite{CMS:2016kce} and reinterpreted results (in double parentheses) for other types of particles. 
Table taken from \cite{Acharya:2020uwc}.\\[-8pt]}
\centering
\begin{tabular}{c|c|c}
        & MoEDAL & (CMS) \\\hline
         Scalar singlet & 160 & ((320))\\
         Fermion singlet & 650 & (680)\\
         Scalar triplet & 340 & ((590))\\
         Fermion triplet & 1130 & ((900))
\end{tabular}
\label{tab:prospects:susy:doubly}
\end{minipage}
\vspace{-1em}
\end{table}

\subsection{Testing radiative neutrino mass models}\label{sec:prospects:neutrino}

Neutrino oscillations constitute clear evidence for the BSM physics, since this phenomenon can 
occur only if neutrinos are massive, contrary to Standard Model predictions. It bears no 
surprise that many theoretical new physics models concentrate on explaining 
neutrino masses. One of such models has been proposed in Ref. \cite{R:2020odv}, 
where neutrino masses are generated at the 1-loop level. The model introduces several new 
fields, including scalar $\rm SU(2)_L$ singlet, $S_1$, and triplet, $S_3$, and three pairs of 
vector-like spin-1/2 fermions $(F_i \bar F_i)(i=1,2,3)$ in $\rm SU(2)_L$ doublet 
representations. The hypercharge assignment is 2, 3, 5/2, and $-5/2$ for $S_1$, $S_3$, $F_i$, and $\bar F_i$, respectively. The Lagrangian of the model is given by:
\begin{equation}\label{eq:prospects:neutrino:lagrangian}
\begin{aligned}
\mathcal{L}_{\rm BSM} =\; & \mathcal{L}_{\rm KIN}
 - \Big[
    (h_{ee})_{ij} (e^i_R)^C (e^j_R)^C S_1^\dag 
  + (h_F)_{ij} L_i F_j S_1^\dag 
  + (h_{\bar F})_{ij} L_i \bar{F}_j S_3^\dag 
  + \text{h.c.}
 \Big] + \\
& + \lambda_2 |H|^2 |S_1|^2 
 + \lambda_{3a} |H|^2 |S_3|^2 
- \Big[ \lambda_5 H H S_1 S^\dag_3 + \text{h.c.} \Big] +\\
& + \lambda_4 |S_1|^4 
 + \lambda_{6a} |S_3^\dag S_3|^2
 + \lambda_{6b} |S_3 S_3 S_3^\dag S_3^\dag| 
 + \lambda_7 |S_1|^2 |S_3|^2 .
\end{aligned}
\end{equation}
The experimental constraints on neutrino masses require that the product of $\lambda_5$ and $h_F h_{\bar F}$ couplings is small.

The feature of this model most relevant for MoEDAL is the presence of multiply charged mass 
eigenstates arising from the triplet scalar field, namely \(S^{\pm 4}\), \(S^{\pm 3}\), and \(S^{\pm 2}\). These states are naturally long-lived provided that neither \(\lambda_5\) nor 
the product \(h_F h_{\bar F}\) is large. This makes the model testable at MoEDAL, and a 
dedicated study of this possibility has been presented in Ref. \cite{Hirsch:2021wge}.

Ref. \cite{Hirsch:2021wge} contains prospects to test the model described by Eq.\ \eqref{eq:prospects:neutrino:lagrangian}, including a comprehensive investigation of the 
impact of various couplings on the final sensitivity. In addition, a coloured version of the 
model is discussed in Ref. \cite{Hirsch:2021wge}, which is obtained by promoting new scalar and 
fermionic fields to (anti-)triplets under $\rm SU(3)_C$, and shifting their hypercharges by $-2/3$. In this article, we constrain ourselves to presenting the expected sensitivity of 
MoEDAL to multiply charged long-lived particles in separation from the original model. In 
other words, we treated masses and lifetimes of particles as free parameters, and we 
considered only particle-antiparticle pair creation\footnote{In the full model, mixed production 
is possible, e.g. $S^{3+}$Sec.  can be produced together with $S^{2-}$ via s-channel W boson exchange.}. We used Eq.\ \eqref{eq:prospects:susy:pntd} when calculating MoEDAL's sensitivity.
Unlike previous studies described in Sec.\ \ref{sec:prospects:neutralino} and Sec.\ \ref{sec:prospects:susy}, we included additional production modes via photon-fusion, which we find non-negligible. This resulted in improved limits compared to Tab.\ \ref{tab:prospects:susy:doubly} for doubly 
charged scalars.

Tab.\ \ref{tab:prospects:neutrino:singlet} summarises MoEDAL's expected sensitivity to colour-singlet long-lived scalars and fermions. Results are shown for both Run 3 ($L=30~{\rm fb}^{-1}$) 
and HL-LHC ($L=300~{\rm fb}^{-1}$), and for $N_{\rm sig}=1$ and $N_{\rm sig}=3$. To compare 
MoEDAL's prospects for BSM detection with those of the major LHC experiments, we used limits from 
the ATLAS search for heavy long-lived multi-charged particles \cite{ATLAS:2018imb}, based on $L=36~{\rm fb}^{-1}$. In that study, the collaboration interpreted their results only for spin-1/2 fermions. We therefore estimated the corresponding limits for scalars by taking the cross-section bounds from \cite{ATLAS:2018imb} and assuming that the efficiencies are not strongly 
dependent on spin. For Run 3, the only projection available at the time of writing was for 
fermions \cite{Jager:2018ecz}, with an estimated reach of about 1500 GeV.

The conclusion from Tab.\ \ref{tab:prospects:neutrino:singlet} is that, by the end of Run 3, 
MoEDAL is expected to provide weaker sensitivity to the considered BSM particles than the current 
constraints from the major LHC experiments, even when requiring only a single detected particle. 
At the end of the HL-LHC, MoEDAL’s sensitivity to $S^{\pm 2}$ would remain below the ATLAS limit 
obtained with $L=36~{\rm fb}^{-1}$. However, for $S^{\pm 3}$ and $S^{\pm 4}$, MoEDAL would exceed 
that sensitivity. For $F^{\pm 3}$, MoEDAL’s HL-LHC reach would be slightly better than the ATLAS 
projection for the end of Run 3, but only if we use $\rm N_{sig}=1$ to set the limit. Overall, this indicates that while MoEDAL cannot compete with 
the major experiments in setting the strongest mass limits, its fundamentally different detection 
technique provides complementary and independent constraints.

\begin{table}[t]
    \caption{\small Model-independent mass limits on long-lived multiply charged colour-singlet BSM particles. 
    The first column contains limits for fermions from Ref. \cite{ATLAS:2018imb} for $L=36~{\rm fb}^{-1}$, and our recast for scalar particles (in double parentheses). 
    The second column contains the projection for fermions for Run 3 $L=300~{\rm fb}^{-1}$, taken from \cite{Jager:2018ecz}. The last two columns contain the expected sensitivity of MoEDAL for $L=30~{\rm fb}^{-1}$ and $L=300~{\rm fb}^{-1}$, respectively, for $\rm N_{sig}=3$ ($\rm N_{sig}=1$). All numbers are in GeV. Table taken from \cite{Hirsch:2021wge}. \\[-5pt]}
     \centering
    \begin{tabular}{c|c|c|c|c}
        & \makecell{current HSCP bound \\ ($36\,\mathrm{fb}^{-1}$)}
        & \makecell{HSCP (Run 3) \\ ($300\,\mathrm{fb}^{-1}$)}
        & \makecell{MoEDAL (Run 3) \\ ($30\,\mathrm{fb}^{-1}$)}
        & \makecell{MoEDAL (HL-LHC) \\ ($300\,\mathrm{fb}^{-1}$)} \\ \hline
        $S^{\pm 2}$ & ((650)) & - & 190 (290) & 400 (600) \\
        $S^{\pm 3}$ & ((780)) & - & 430 (610) & 850 (1100) \\
        $S^{\pm 4}$ & ((920)) & - & 700 (960) & 1200 (1430) \\
        $F^{\pm 3}$ & 1130 & 1500 & 800 (1030) & 1300 (1550)
    \end{tabular}
    \label{tab:prospects:neutrino:singlet}
    \vspace{-1em}
\end{table}

Tab.\ \ref{tab:prospects:neutrino:triplet} contains MoEDAL's sensitivity to coloured scalars: $\tilde S^{\pm 4/3}$, $\tilde S^{\pm 7/3}$, and $\tilde S^{\pm 10/3}$. The limits are significantly stronger than the 
limits in Tab.\ \ref{tab:prospects:neutrino:triplet}, mainly due to larger production cross-sections. To compare MoEDAL's discovery prospects with ATLAS and CMS, we again refer to Ref. \cite{Jager:2018ecz}, where 7 and 8 TeV CMS results \cite{CMS:2013czn} were rescaled to 13 TeV 
and $L=36~{\rm fb}^{-1}$. Projections for Run 3 ($L=300~{\rm fb}^{-1}$) were also taken from Ref. \cite{Jager:2018ecz}. The result shown in Tab.\ \ref{tab:prospects:neutrino:triplet} reveals 
that MoEDAL can offer a comparable sensitivity to ATLAS/CMS only for $\tilde S^{\pm 10/3}$ when both 
experiments use $L=300~{\rm fb}^{-1}$ data set size. However, MoEDAL will collect this amount of 
data only at the end of the HL-LHC phase, while ATLAS and CMS will achieve it at the end of Run 
3.

\begin{table}[t]
    \caption{ \small
    Model-independent mass limits on long-lived multiply charged colour-(anti)triplet scalar particles. The first column contains limits (in double parentheses) estimated in Ref. \cite{Jager:2018ecz}, by scaling up 7 and 8 TeV CMS results \cite{CMS:2013czn} to 13 TeV and $L=30~{\rm fb}^{-1}$. The second column contains projections at the end of Run 3 ($L=300~{\rm fb}^{-1}$), taken from \cite{Jager:2018ecz}. The last two columns contain the expected sensitivity of MoEDAL for $L=30~{\rm fb}^{-1}$ and $L=300~{\rm fb}^{-1}$, respectively, for $\rm N_{sig}=3$ ($\rm N_{sig}=1$). All numbers are in GeV. Table taken from \cite{Hirsch:2021wge}. \\[-5pt]
    }
    \centering
    \begin{tabular}{c|c|c|c|c}
        & \makecell{current HSCP bound \\ ($36\,\mathrm{fb}^{-1}$)}
        & \makecell{HSCP (Run 3) \\ ($300\,\mathrm{fb}^{-1}$)}
        & \makecell{MoEDAL (Run 3) \\ ($30\,\mathrm{fb}^{-1}$)}
        & \makecell{MoEDAL (HL-LHC) \\ ($300\,\mathrm{fb}^{-1}$)} \\ \hline
        $\tilde S^{\pm 4/3}$ & ((1450)) & 1700 & 880 (1050) & 1250 (1400) \\
        $\tilde S^{\pm 7/3}$ & ((1480)) & 1730 & 1080 (1250) & 1450 (1650) \\
        $\tilde S^{\pm 10/3}$ & ((1510)) & 1790 & 1200 (1400) & 1600 (1800) \\
    \end{tabular}
    \label{tab:prospects:neutrino:triplet}
    \vspace{-1em}
\end{table}

\FloatBarrier
\subsection{Searches for multiply charged BSM particles}\label{sec:prospects:multicharged}

The final paper in the series, Ref.~\cite{Altakach:2022hgn}, was focused on simplified BSM scenarios featuring 
multiply charged particles. The study considered both scalar and spin-1/2 fermionic states, taken either as singlets 
or triplets under $\mathrm{SU(3)_C}$, with electric charges spanning from $\pm 1e$ up to $\pm 8e$. All new fields 
were assumed to be $\mathrm{SU(2)_L}$ singlets. In total, discovery prospects for 32 distinct classes of BSM 
particles at the LHC were systematically evaluated.

A special emphasis was put on an accurate production cross-section estimation. At the time of 
publication of Ref. \cite{Altakach:2022hgn}, experimental searches for multiply charged BSM 
particles, e.g. \cite{ATLAS:2018imb}, were assuming Drell-Yan production of BSM particles. While this is a 
justified assumption for particles with $q=\pm 1 e$ or $q = \pm 2 e$, it lead to significant 
underestimation of production cross-section for particles with larger electric charges, for which photon fusion (and gluon-photon fusion for coloured particles) becomes non-negligible.

To estimate discovery prospects for multiply charged long-lived BSM particles at the LHC, three search strategies 
were considered: (1) anomalous track searches (i.e. large $dE/dx$ analyses) at CMS \cite{CMS:2013czn, CMS:2016kce} and ATLAS \cite{ATLAS:2019gqq}, (2) searches 
for charged LLPs at MoEDAL, (3) diphoton resonance searches \cite{ATLAS:2021uiz}. The first two types of searches assume a production 
of a particle-antiparticle pair that propagates through the detector without decaying. The latter search strategy 
considers the formation of a bound state due to attractive electromagnetic and/or strong force, and its decay into a 
pair of photons, $pp \to \mathcal{B}\to \gamma \gamma$.

\begin{figure}[h]
    \centering
    \begin{tabular}{cc}
        \includegraphics[width=0.48\textwidth]{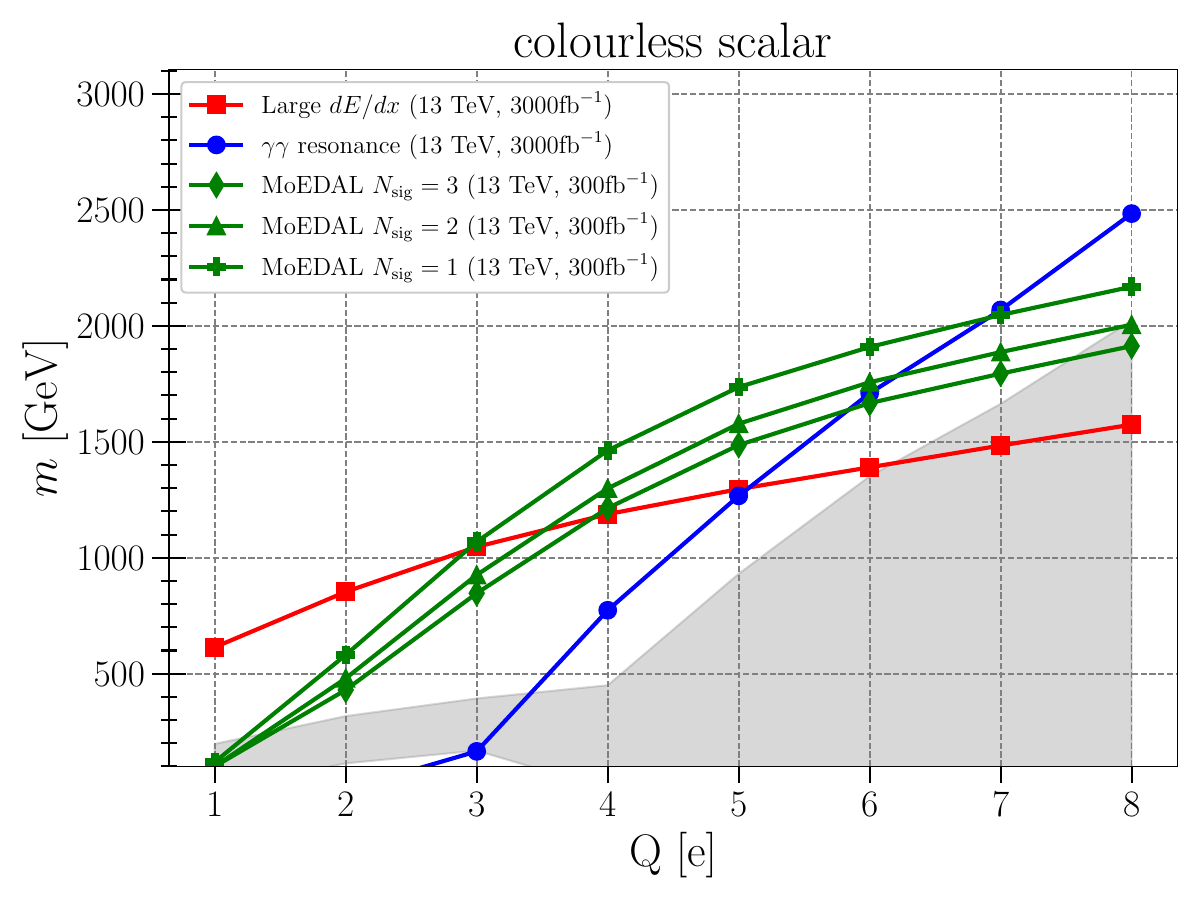} &
        \includegraphics[width=0.48\textwidth]{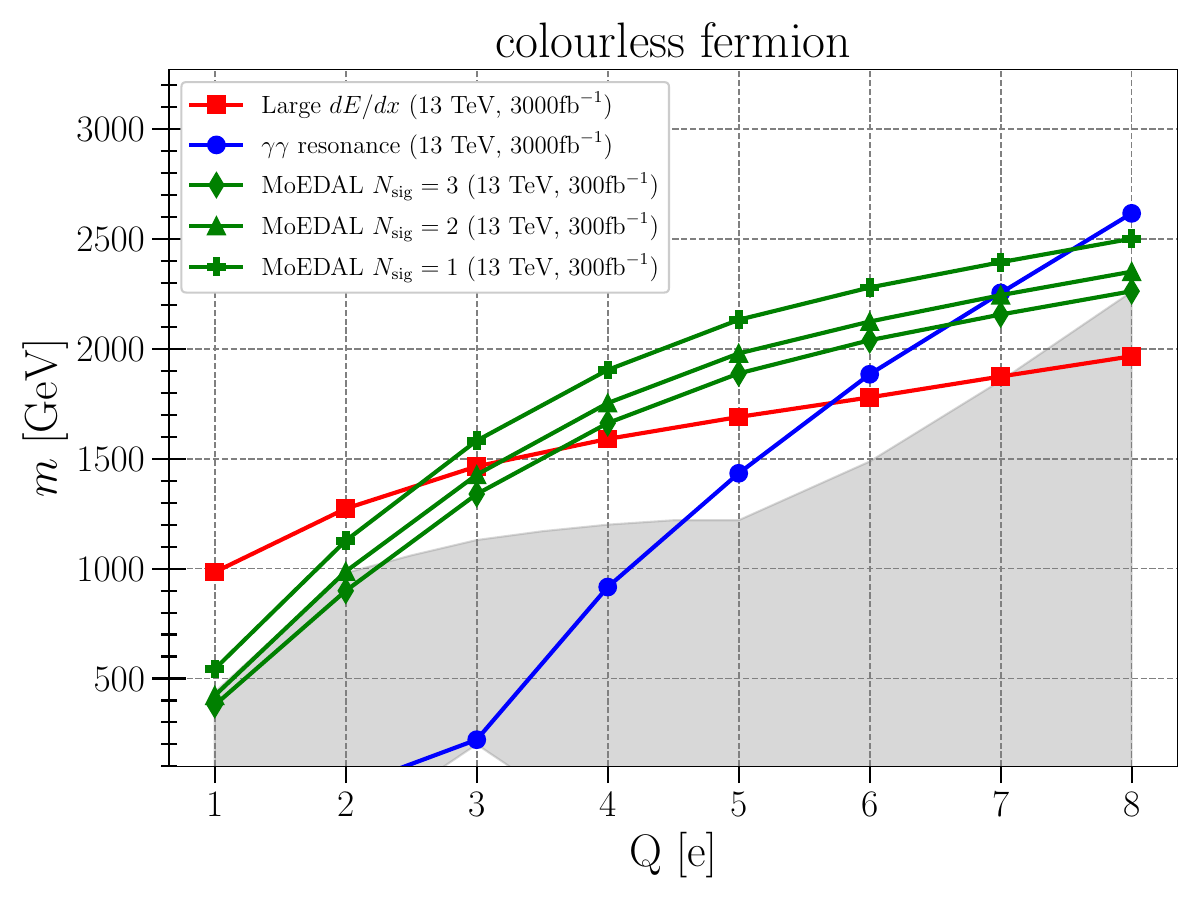} \\[6pt]
        \includegraphics[width=0.48\textwidth]{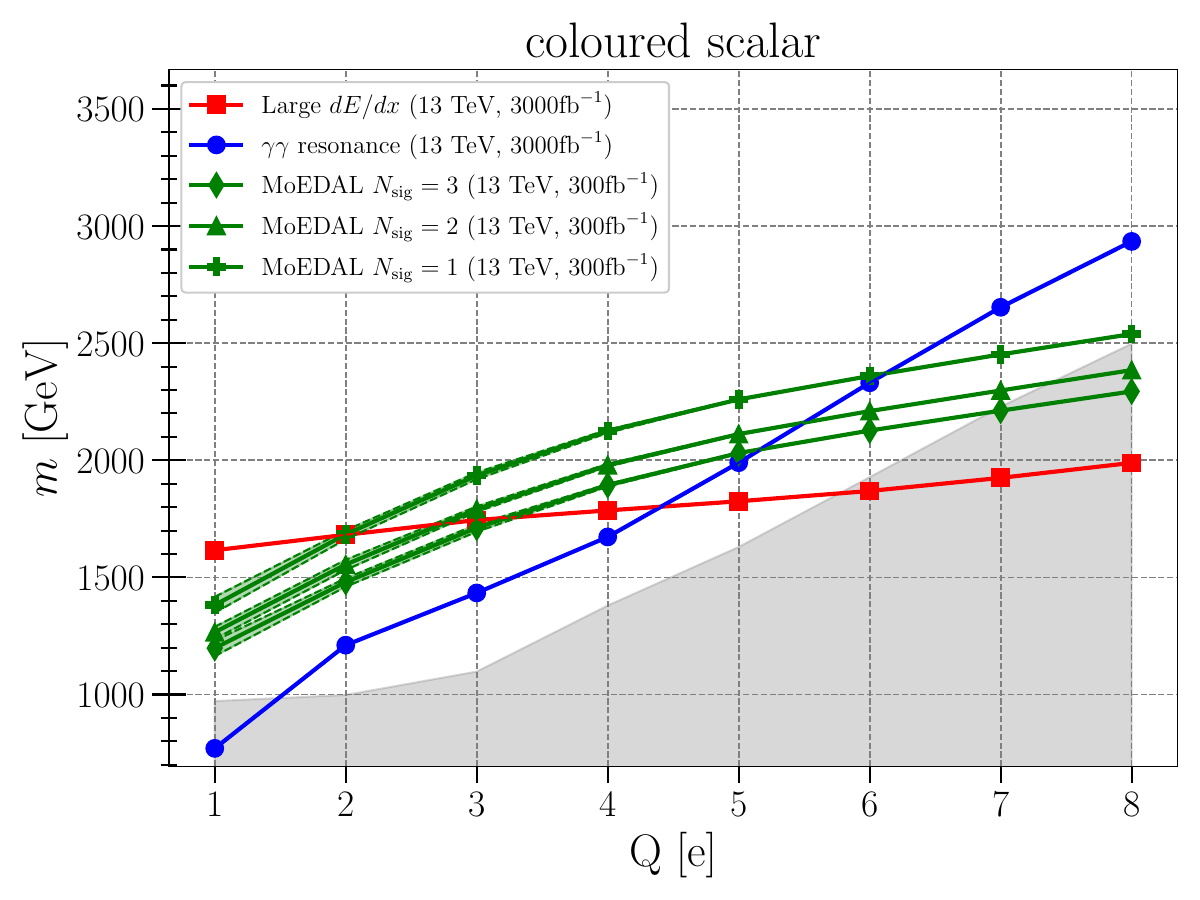} &
        \includegraphics[width=0.48\textwidth]{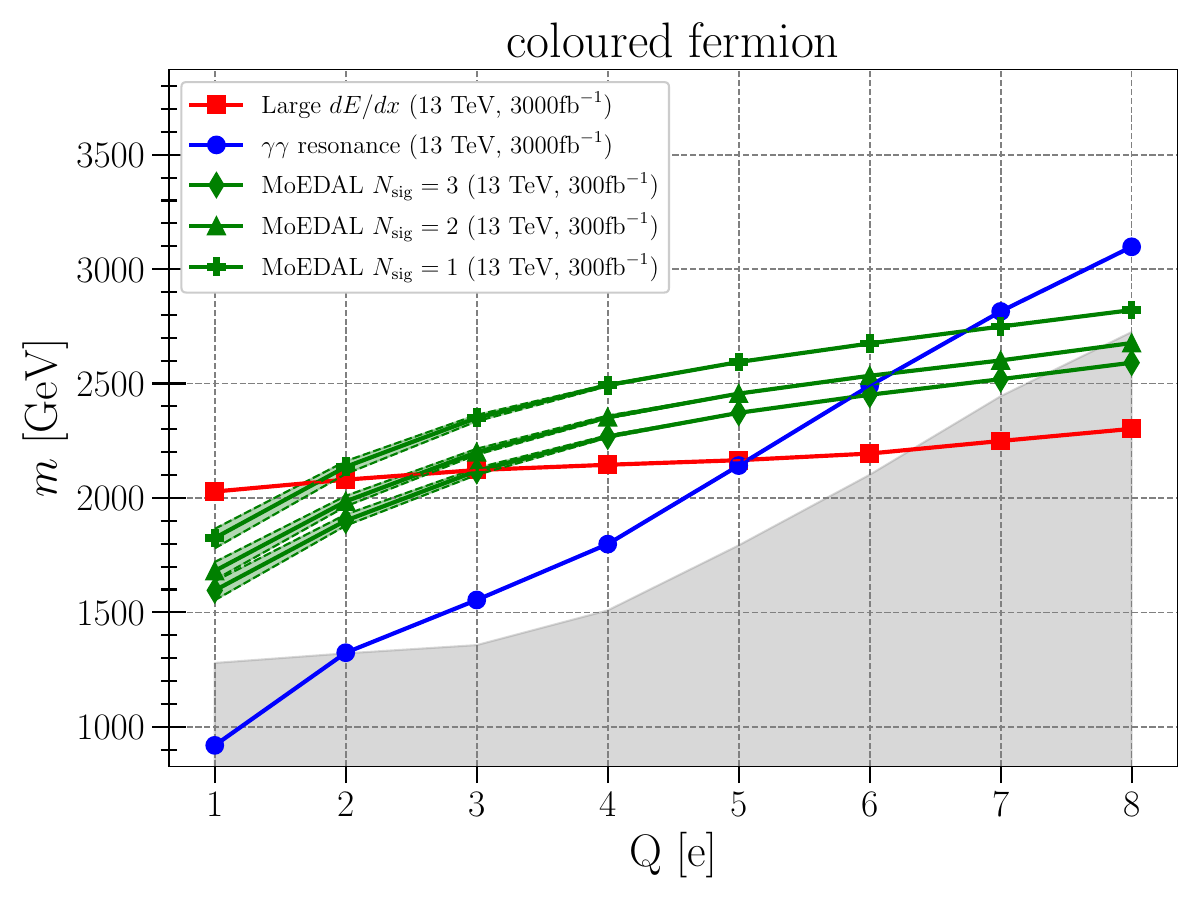}
    \end{tabular}
    \caption{Expected sensitivity of LHC searches for multiply charged long-lived particles at the end of HL-LHC. Results are for colour-singlet scalars (top left), colour-singlet fermions (top right), colour-triplet scalars (bottom left), and colour-triplet fermions (bottom right). Figure taken from \cite{Altakach:2022hgn}.
    \vspace{-1em}
    }
    \label{fig:prospects:multicharged:results}
\end{figure}

Fig. \ref{fig:prospects:multicharged:results} contains the expected sensitivity of the three search strategies to 32 
types of BSM LLPs studied. The results assume the integrated luminosity of $3~{\rm ab^{-1}}$ for ATLAS/CMS and $300~{\rm fb^{-1}}$ for MoEDAL, which corresponds to the end of the HL-LHC phase. Red curves represent the 
expected sensitivity of the large $dE/dx$ searches, obtained by following the recasting procedure in Ref. \cite{Jager:2018ecz} applied to the CMS results in Ref. \cite{CMS:2016kce}. These searches provide the 
strongest limits for particles with small electric charges, i.e. $|q| \leqslant 3 e$. Their sensitivity 
slightly increases with the magnitude of the charge. Blue curves correspond to recasting the diphoton 
resonance searches by ATLAS in Ref. \cite{ATLAS:2021uiz}. These searches provide no (or very weak) limits 
for particles with small electric charges, however, their sensitivity grows rapidly with the magnitude of 
charge, and makes them the most relevant for large charges, i.e. $|q| \geqslant 7e$. Finally, Fig. \ref{fig:prospects:multicharged:results} contains three green curves (with uncertainty bands for coloured 
particles due to hadronisation) representing MoEDAL's sensitivity, calculated with Eq.\ \eqref{eq:prospects:susy:pntd}, for $N_{\rm sig}=1,2,3$ particles 
detected. These limits grow with the magnitude of the electric charge. Interestingly, for intermediate 
charge magnitudes, i.e. $ 3e \leqslant |q|\leqslant 6e$, MoEDAL is able to provide better limits than 
major LHC experiments, despite $\sim 10$ lower luminosity. This is due to MoEDAL being background-free. 
For ATLAS and CMS, both background and signal scale up with increased luminosity, while for MoEDAL, only 
the signal grows; the background remains $\ll 1$. Finally, Fig. \ref{fig:prospects:multicharged:results} provides a comparison with current experimental constraints by depicting them as grey-shaded areas.

\section{Conclusions}\label{sec:conclusions}

Searches for charged long-lived BSM particles constitute a rapidly growing and interesting branch of 
collider studies. In this article, we have summarised prospects for detecting various kinds of charged 
LLPs in future years, focusing on the MoEDAL experiment. We have estimated MoEDAL's sensitivity to 
particles in the MSSM, the radiative neutrino mass model, and various simplified models. We compared our 
findings to relevant constraints provided by ATLAS and CMS experiments. 

We found that, in general scenarios with singly or doubly charged particles, MoEDAL offers lower sensitivity than major LHC experiments, mainly due to the lower luminosity available at the IP8. However, 
in some special cases, e.g. involving intermediate long-lived neutral particles like the one described in 
Sec.\ \ref{sec:prospects:neutralino}, MoEDAL can be more sensitive to new particles.

Due to its special detector design, MoEDAL achieves sensitivity comparable to ATLAS and CMS for 
particles with electric charges in the range $3e \lesssim |q| \lesssim 6e$. In the High-Luminosity LHC 
phase, MoEDAL is expected to be competitive with the general-purpose experiments, primarily due to 
its essentially background-free design.

\backmatter

{\small
\bmhead{Acknowledgements} 
This publication is co-funded by/ has received funding from/ the European Union’s
Horizon Europe research and innovation program under the Marie Sklodowska-Curie
COFUND Postdoctoral Programme grant agreement No.101081355-SMASH.
The operation (SMASH project) is co-funded by the Republic of Slovenia and the
European Union from the European Regional Development Fund.

\bmhead{Statement}
Co-funded by the European Union. Views and opinions expressed are, however, those of the author(s) only and do not necessarily reflect
those of the European Union or European Research Executive Agency. Neither the European Union nor the granting authority can be held responsible for them.
}
\newpage
\bibliography{sn-bibliography}

\end{document}